\documentclass{AIAA}
\usepackage{amsfonts}
\usepackage{amsmath}
\usepackage{amssymb}
\usepackage{graphicx}
\usepackage{amsfonts}
\usepackage{url}

\begin{document}

\title{Terminal Iterative Learning Control for Autonomous Aerial Refueling
under Aerodynamic Disturbances}

\author{Xunhua Dai\footnote{Ph.D. Candidate, School of Automation Science and Electrical Engineering;
dai@buaa.edu.cn.}, Quan Quan\footnote{Associate Professor, School of Automation Science and Electrical Engineering;
qq\_buaa@buaa.edu.cn.}, Jinrui Ren\footnote{Ph.D. Candidate, School of Automation Science and Electrical Engineering;
renjinrui@buaa.edu.cn.}, Zhiyu Xi\footnote{Associate Professor, School of Automation Science and Electrical Engineering;
z.xi@buaa.edu.cn.} and Kai-Yuan Cai\footnote{Professor, School of Automation Science and Electrical Engineering;
kycai@buaa.edu.cn.}}
\affiliation{Beihang University, 100191 Beijing, People's Republic of China}

\maketitle

\section*{Nomenclature}

\begin{tabular}{@{}lcl}
${\Delta\mathbf{F}_{0}}$  & =  & Bow wave disturbance force \tabularnewline
$\Delta{\mathbf{p}}_{\text{dr}}^{\text{e}}$  & =  & Drogue position offsets from equilibrium position \tabularnewline
$\Delta{\mathbf{p}}_{\text{dr/pr}}$  & =  & Position error between drogue and probe \tabularnewline
$\Delta{\mathbf{p}}_{\text{dr/pr}}\left(T\right)$  & =  & Terminal position error \tabularnewline
$\Delta R_{\text{dr/pr}}$  & =  & Radial error between hose and drogue \tabularnewline
$F_{\text{T}}$  & =  & Tanker joint frame \tabularnewline
{$\mathbf{F}$}$_{\text{r}}$, {$\mathbf{F}$}$_{\text{hd}}$  & =  & Disturbance forces on receiver and hose-drogue \tabularnewline
{$\mathbf{F}$}$_{\text{bow}}$  & =  & Disturbance force from bow wave effect \tabularnewline
${\mathbf{p}}_{\text{dr}}\left(t\right)$, ${\mathbf{p}}_{\text{pr}}\left(t\right)$  & =  & Current positions of drogue and probe \tabularnewline
${\mathbf{p}}_{\text{dr}}\left(T\right)$, ${\mathbf{p}}_{\text{pr}}\left(T\right)$  & =  & Terminal positions of drogue and probe \tabularnewline
${\mathbf{p}}_{\text{dr}}^{\text{e0}}$  & =  & Drogue initial equilibrium position \tabularnewline
$
\mathbb{R}
$, $
\mathbb{R}
_{+}$  & =  & Real number set and positive real number set \tabularnewline
$R_{\text{C}}$  & =  & Threshold radius for a successful docking attempt \tabularnewline
$T$  & =  & Terminal time of a docking attempt \tabularnewline
$\mathbf{\hat{u}}_{\text{pr}}$  & =  & Reference trajectory for autopilot \tabularnewline
\end{tabular}


\section{Introduction}

Aerial refueling has demonstrated significant benefits to aviation
by extending the range and endurance of aircraft \cite{Nalepka-2005-1}.
The development of autonomous aerial refueling (AAR) techniques for
unmanned aerial vehicles (UAVs) makes new missions and capabilities
possible \cite{Dibley-2007-2}, like the ability for long range or
long time flight. As the most widely used aerial refueling method,
the probe-drogue refueling (PDR) system is considered to be more flexible
and compact than other refueling systems. However, a drawback of PDR
is that the drogue is passive and susceptible to aerodynamic disturbances
\cite{AAR-2014}. Therefore, it is difficult to design an AAR system
to control the probe on the receiver to capture the moving drogue
within centimeter level in the docking stage.

It used to be thought that the aerodynamic disturbances in the aerial
refueling mainly include the tanker vortex, wind gust, and atmospheric
turbulence. According to NASA Autonomous Aerial Refueling Demonstration
(AARD) project \cite{Dibley-2007-2}, the forebody flow field of the
receiver may also significantly affect the docking control of AAR,
which is called ``the bow wave effect'' \cite{Bhandari-2013-8}.
As a result, the modeling and simulation methods for the bow wave
effect were studied in our previous works \cite{dai2016modeling,wei2016drogue}.
Since the obtained mathematical models are somewhat complex and there
may be some uncertain factors in practice, this paper aims to use
a model-free method to compensate for the docking error caused by
aerodynamic disturbances including the bow wave effect.

Most of the existing studies on AAR docking control do not consider
the bow wave effect. In \cite{tandale2006trajectory,zhu2016vision,liu2017modeling},
the drogue is assumed to be relatively static (or oscillates around
the equilibrium) and not affected by the flow field of the receiver
forebody. However, in practice, the receiver aircraft is affected
by aerodynamic disturbances, and the drogue is affected by both the
wind disturbances and the receiver forebody bow wave. As a major difficulty
in the control of AAR, the aerodynamic disturbances, especially the
bow wave effect, attract increasing attention in these years. In \cite{Vortex-1,lee2013estimation},
the wind effects from the tanker vortex, the wind gust, and the atmospheric
turbulence are analyzed, and in \cite{dai2016modeling,wei2016drogue,Khan-2014-9},
the modeling and simulation methods for the receiver forebody bow
wave effect are studied, but no control methods are proposed. In \cite{Bhandari-2013-8},
simulations show that the bow wave effect can be compensated by adding
an offset value to the reference trajectory, but the method for obtaining
the offset value is not given.

Since the accurate mathematical models for the aerodynamic disturbances
are usually difficult to obtain \cite{dai2016modeling}, iterative
learning control (ILC) is a possible choice for the docking control
of AAR. According to \cite{bristow2006survey}, the ILC is a model-free
control method which can improve the performance of a system by learning
from the previous repetitive executions or iterations. ILC methods
have been proved to be effective to solve the control problems for
complex systems with no need for the exact mathematical model \cite{ahn2007iterative}.
For an actual AAR system, the relative position between the probe
and the drogue is usually measured by vision localization methods
\cite{valasek2005vision} whose measurement precision depends on the
relative distance (higher precision in a closer distance). Therefore,
compared with the trajectory data, the terminal positions of the probe
and the drogue are usually easier to measure in practice. As a result,
terminal iterative learning control (TILC) methods are suitable for
AAR systems because TILC methods need only the terminal states or
outputs instead of the whole trajectories \cite{chen1999iterative,chi2014improved}.

This paper studies the model of the probe-drogue aerial refueling
system under aerodynamic disturbances, and proposes a docking control
method based on TILC to compensate for the docking errors caused by
aerodynamic disturbances. In the ATP-56(B) issued by NATO \cite{NATO-2004-3},
chasing the drogue directly is identified as a dangerous operation
which may cause the overcontrol of the receiver. Therefore, the proposed
TILC controller is designed by imitating the docking operations of
human pilots to predict the terminal position of the drogue with an
offset to compensate for the docking errors caused by aerodynamic
disturbances. The designed controller works as an additional unit
for the trajectory generation of the original autopilot system. Simulations
based on our previously published MATLAB/SIMULINK environment \cite{dai2016modeling,wei2016drogue}
show that the proposed control method has a fast learning speed to
achieve a successful docking control under aerodynamic disturbances
including the bow wave effect.

The paper is organized as follows. \textit{Section \ref{Sec-2}} gives
comprehensive problem description and model analysis of a PDR system.
\textit{Section \ref{Sec-3}} describes the details of the TILC controller
design and the convergence analysis. \textit{Section \ref{Sec-4}}
gives simulations with the proposed TILC method. In the end, \textit{Section
\ref{Sec-5}} presents the conclusions.

\section{Problem Formulation}

\label{Sec-2}

\subsection{Frames and Notations}

\begin{figure}[tbh]
\centering \includegraphics[width=0.45\textwidth]{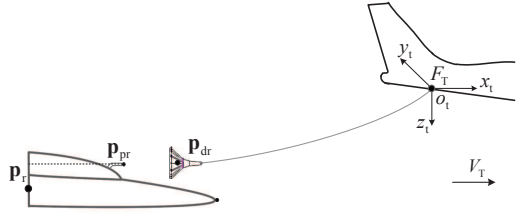}
\caption{Simplified schematic diagram of PDR systems.}
\label{Fig_OverViewAAR}
\end{figure}

Since the tanker moves at a uniform speed in a straight and level
line during the docking stage of AAR, a frame fixed to the tanker
body can be treated as an inertial reference frame to describe the
relative motion between the receiver and the drogue. As shown in Fig.~\ref{Fig_OverViewAAR},
a tanker joint frame $F_{\text{T}}$ is defined with the origin $O_{\text{t}}$
fixed to the joint between the tanker body and the hose. $F_{\text{T}}$
is a right-handed coordinate system, whose $x_{\text{t}}$ horizontally
points to the flight direction of the tanker, $z_{\text{t}}$ vertically
points to the ground, and $y_{\text{t}}$ points to the right. For
simplicity, the following rules are defined:

(i) All position or state vectors are defined under the tanker joint
frame $F_{\text{T}}$, unless explicitly stated.

(ii) The drogue position vector is expressed as ${\mathbf{p}}_{\text{dr}}\triangleq\left[x_{\text{dr}}\;y_{\text{dr}}\;z_{\text{dr}}\right]^{\text{T}}$,
and the probe position vector is ${\mathbf{p}}_{\text{pr}}\triangleq\left[x_{\text{pr}}\;y_{\text{pr}}\;z_{\text{pr}}\right]^{\text{T}}$.
In a similar way, the position error between the probe and the drogue
is expressed as
\begin{equation}
\Delta{\mathbf{p}}_{\text{dr/pr}}\left(t\right)\triangleq{\mathbf{p}}_{\text{dr}}\left(t\right)-{\mathbf{p}}_{\text{pr}}\left(t\right)\label{Eq-1-1}
\end{equation}
whose decomposition form is represented by $\Delta{\mathbf{p}}_{\text{dr/pr}}\triangleq[\Delta x_{\text{dr/pr}}\;\Delta y_{\text{dr/pr}}\;\Delta z_{\text{dr/pr}}]^{\text{T}}$
.

(iii) One docking attempt ends at the terminal time $T\in
\mathbb{R}
_{+}$ when the probe contacts with the central plane of the drogue ($\Delta x_{\text{dr/pr}}=0$)
for the first time, which is defined as
\begin{equation}
T=\min_{t}\left\{ \Delta x_{\text{dr/pr}}\left(t\right)=0\right\} .\label{Eq-1-2}
\end{equation}
The value at time $t=T$ is called the terminal value. For example,
${\mathbf{p}}_{\text{dr}}\left(T\right)$ is the terminal position
of the drogue and $\Delta{\mathbf{p}}_{\text{dr/pr}}\left(T\right)$
is the terminal position error.

(iv) The value in the $k^{\text{th}}$ docking attempt is marked by
a right superscript. For example, ${\mathbf{p}}_{\text{dr}}^{\left(k\right)}$
denotes the drogue position ${\mathbf{p}}_{\text{dr}}$ in the $k^{\text{th}}$
docking attempt, $T^{\left(k\right)}$ denotes the $k^{\text{th}}$
terminal time, and ${\mathbf{p}}_{\text{dr}}^{\left(k\right)}\left(T^{\left(k\right)}\right)$
denotes the $k^{\text{th}}$ terminal position of the drogue.

\subsection{System Overview}

\begin{figure}[tbh]
\centering \includegraphics[width=0.8\textwidth]{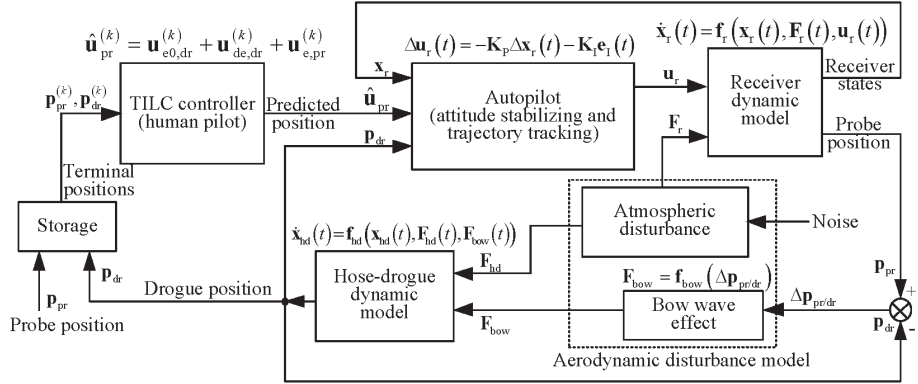}
\caption{Overall structure of the AAR system.}
\label{Fig-InnerCtrl}
\end{figure}

The overall structure of the AAR system proposed in this paper is
shown in Fig.~\ref{Fig-InnerCtrl}, where the whole AAR system is
divided into two parts: the mathematical model and the control system.
The AAR Mathematical model contains three components: the aerodynamic
disturbance model, the hose-drogue dynamic model, and the receiver
dynamic model; the control system contains two components: the autopilot
and the TILC controller. The autopilot focuses on stabilizing the
aircraft attitude and tracking the given reference trajectory, and
the TILC controller works as a human pilot that learns from historical
experience and sends trajectory commands to the autopilot. This paper
focuses on the design of the TILC controller.

\subsection{Mathematical Model}

\subsubsection{Aerodynamic Disturbance Model}

The aerodynamic disturbances will change the flow field around the
receiver and the drogue, then produce disturbance forces on them to
affect their relative motions. There are mainly two sources of aerodynamic
disturbances: one is from the atmospheric environment such as the
tanker vortex, the wind gust and the atmospheric turbulence \cite{Vortex-1};
the other is from the bow wave flow field of the receiver forebody.
In an AAR system, the receiver mainly suffers the atmospheric disturbance
force {$\mathbf{F}$}$_{\text{r}}\in
\mathbb{R}
^{3}$, while the hose-drogue suffers both the atmospheric disturbance force
{$\mathbf{F}$}${_{\text{hd}}}\in
\mathbb{R}
^{3}$ and the bow wave disturbance force {$\mathbf{F}$}$_{\text{bow}}\in
\mathbb{R}
^{3}$.

The modeling and simulation methods for {$\mathbf{F}$}$_{\text{r}}$
and {$\mathbf{F}$}${_{\text{hd}}}$ have been well studied in the
existing literature, where the detailed mathematical expression for
{$\mathbf{F}$}$_{\text{r}}$ can be found in \cite{Vortex-1},
the detailed mathematical expression for {$\mathbf{F}$}${_{\text{hd}}}$
can be found in \cite{dai2016modeling}\cite{Vortex-1}. The bow wave
disturbance force {$\mathbf{F}$}$_{\text{bow}}$, according to
\cite{dai2016modeling}, is determined by the position error between
the drogue and the probe $\Delta{\mathbf{p}}_{\text{dr/pr}}$, which
can be expressed as
\begin{equation}
{\mathbf{F}}_{\text{bow}}{=\mathbf{f}}_{\text{bow}}\left(\Delta{\mathbf{p}}_{\text{dr/pr}}\right)\label{Eq-1-3}
\end{equation}
where ${\mathbf{f}}_{\text{bow}}\left(\cdot\right)$ is the bow wave
effect function whose expression can be obtained by the method proposed
in \cite{dai2016modeling}.

Among these disturbances, {$\mathbf{F}$}$_{\text{r}}$ and {$\mathbf{F}$}${_{\text{hd}}}$
are independent of the states of the AAR system, and the corresponding
control methods are mature; {$\mathbf{F}$}$_{\text{bow}}$ is strongly
coupled with the system output $\Delta{\mathbf{p}}_{\text{dr/pr}}$,
the control strategy for which is challenging and still lacking. Therefore,
this paper puts more effort on the control of the bow wave effect.

\subsubsection{Hose-drogue Model}

The soft hose can be modeled by a finite number of cylinder-shaped
rigid links based on the finite-element theory \cite{hose-link-model}.
Then, the hose-drogue dynamic equation can be written as
\begin{equation}
\left\{ \begin{array}{l}
{\mathbf{{\dot{x}}}}_{\text{hd}}\left(t\right)={\mathbf{f}{_{\text{hd}}}}\left({{\mathbf{x}_{\text{hd}}}\left(t\right),}\mathbf{F}_{\text{hd}}\left(t\right),\mathbf{F}_{\text{bow}}\left(t\right)\right)\\
{\mathbf{p}_{\text{dr}}}\left(t\right)={\mathbf{g}{_{\text{hd}}}}\left({\mathbf{x}_{\text{hd}}}\left(t\right)\right)
\end{array}\right.\label{Eq_hoseDorgueDyn}
\end{equation}
where ${\mathbf{f}{_{\text{hd}}}}\left(\cdot\right)$ is a nonlinear
vector function{,} $\mathbf{x}_{\text{hd}}$ is the hose-drogue
state vector, and $\mathbf{F}_{\text{hd}}\left(t\right)$ and $\mathbf{F}_{\text{bow}}\left(t\right)$
are the disturbance forces acting on the drogue. The dimensions of
$\mathbf{x}_{\text{hd}}$ and ${\mathbf{f}{_{\text{hd}}}}\left(\cdot\right)$
depend on the number of the links that the hose is divided into.

The most concerned value in the TILC method is the terminal position
of the drogue. Therefore, it is necessary to study the terminal state
of the hose-drogue system (\ref{Eq_hoseDorgueDyn}). According to
\cite{wei2016drogue}, when there is no random disturbance, the drogue
will eventually settle at an equilibrium position marked as ${\mathbf{p}}_{\text{dr}}^{\text{e0}}$.
Then, under the bow wave effect, the drogue will be pushed to a new
terminal position ${\mathbf{p}}_{\text{dr}}\left(T\right)$. The drogue
position offset $\Delta{\mathbf{p}}_{\text{dr}}^{\text{e}}\in
\mathbb{R}
^{3}$ is defined as
\begin{equation}
\Delta{\mathbf{p}}_{\text{dr}}^{\text{e}}={\mathbf{p}}_{\text{dr}}\left(T\right)-{\mathbf{p}}_{\text{dr}}^{\text{e0}}\label{Eq-wbow0}
\end{equation}
where $\Delta{\mathbf{p}}_{\text{dr}}^{\text{e}}$ is further determined
by the strength of terminal bow wave disturbance force {$\mathbf{F}$}$_{\text{bow}}\left(T\right)$
as
\begin{equation}
\Delta{\mathbf{p}}_{\text{dr}}^{\text{e}}={\mathbf{f}_{\text{dr}}}\left({\mathbf{F}}_{\text{bow}}\left(T\right)\right).\label{Eq-wbow1}
\end{equation}
Then, substituting Eq.~(\ref{Eq-1-3}) into Eq.~(\ref{Eq-wbow1})
yields
\begin{equation}
\Delta{\mathbf{p}}_{\text{dr}}^{\text{e}}={\mathbf{f}_{\text{dr}}}\left({\mathbf{f}}_{\text{bow}}\left(\Delta{\mathbf{p}}_{\text{dr/pr}}\left(T\right)\right)\right)\triangleq{\mathbf{\bar{f}}}_{\text{dr}}\left(\Delta{\mathbf{p}}_{\text{dr/pr}}\left(T\right)\right).\label{Eq_hoseDrDyn2}
\end{equation}
Noticing that $\Delta{\mathbf{p}}_{\text{dr/pr}}\left(T\right)\approx\mathbf{0}$,
the Taylor Expansion can be applied to Eq.~(\ref{Eq_hoseDrDyn2}),
which results in
\begin{equation}
\Delta{\mathbf{p}}_{\text{dr}}^{e}\approx\mathbf{m}_{0}+\mathbf{M}_{1}\cdot\Delta{\mathbf{p}}_{\text{dr/pr}}\left(T\right)\label{Eq_hoseDrDyn2-1}
\end{equation}
where
\begin{equation}
\mathbf{m}_{0}\triangleq{\mathbf{\bar{f}}}_{\text{dr}}\left(\mathbf{0}\right),\;\mathbf{M}_{1}\triangleq\left.\frac{\partial{\mathbf{\bar{f}}}_{\text{dr}}\left(\mathbf{x}\right)}{\partial\mathbf{x}}\right\vert _{\mathbf{x=0}}.\label{Eq_hoseDrDyn2-2}
\end{equation}
In practice, the drogue is sensitive to the aerodynamic disturbances,
and the actual terminal position of the drogue always oscillates around
its stable position. Therefore, a bounded disturbance term $\mathbf{v}_{\text{dr}}\in
\mathbb{R}
^{3}$ should be added to Eq.~(\ref{Eq_hoseDrDyn2}) as
\begin{equation}
\Delta{\mathbf{p}}_{\text{dr}}^{\text{e}}=\mathbf{m}_{0}+\mathbf{M}_{1}\cdot\Delta{\mathbf{p}}_{\text{dr/pr}}\left(T\right)+\mathbf{v}_{\text{dr}}\label{Eq_hoseDrDyn2-3}
\end{equation}
where $\left\Vert \mathbf{v}_{\text{dr}}\right\Vert \leq B_{\text{dr}}$
represents the position fluctuation of the drogue due to random disturbances
such as atmospheric turbulence. According to Eq.~(\ref{Eq_hoseDrDyn2-3}),
there is a functional relationship between the terminal docking error
$\Delta{\mathbf{p}}_{\text{dr/pr}}\left(T\right)$ and the drogue
bow wave offset $\Delta{\mathbf{p}}_{\text{dr}}^{\text{e}}$. Therefore,
it is possible to use TILC methods to compensate for the bow wave
position offset $\Delta{\mathbf{p}}_{\text{dr}}^{\text{e}}$ with
the terminal docking error $\Delta{\mathbf{p}}_{\text{dr/pr}}\left(T\right)$.

The detailed mathematical expression of ${\mathbf{\bar{f}}}_{\text{dr}}\left(\cdot\right)$
can be obtained through methods in \cite{dai2016modeling}, then the
Jacobian matrix $\mathbf{M}_{1}$ can be obtained from Eq.~(\ref{Eq_hoseDrDyn2-2}).
Since ${\mathbf{\bar{f}}}_{\text{dr}}\left(\cdot\right)$ is monotonically
decreasing along each axial direction, for the receiver aircraft with
symmetrical forebody layout, it is easy to verify that $\mathbf{M}_{1}$
is a negative definite matrix.

\subsubsection{Receiver Aircraft Model}

As previously mentioned, in the docking stage, the tanker joint frame
$F_{\text{T}}$ can be simplified as an inertial frame. Under this
situation, the commonly used aircraft modeling methods as presented
in \cite{AirContrl} can be applied to the receiver aircraft with
the following form
\begin{equation}
\left\{ \begin{array}{l}
{\mathbf{{\dot{x}}}}_{\text{r}}\left(t\right)={\mathbf{f}}_{\text{r}}\left({\mathbf{x}}_{\text{r}}{\left(t\right),\mathbf{F}}_{\text{r}}{\left(t\right),}\boldsymbol{u}_{\text{r}}\left(t\right)\right)\\
\mathbf{p}{_{\text{pr}}}\left(t\right)={\mathbf{g}_{\text{pr}}}\left({\mathbf{x}}_{\text{r}}\left(t\right)\right)
\end{array}\right.\label{Eq-rec}
\end{equation}
where ${\mathbf{f}}_{\text{r}}\left(\cdot\right)$ is a nonlinear
function, ${\mathbf{x}}_{\text{r}}$ is the state of the receiver
and $\mathbf{u}_{\text{r}}$ is the control input of the receiver
aircraft.

Since the nonlinear model (\ref{Eq-rec}) is too complex for controller
design, a linearization method \cite{AirContrl} is applied to Eq.~(\ref{Eq-rec})
to simplify the receiver dynamic model. Assume the receiver equilibrium
state is $\mathbf{x}_{\text{r0}}$ and the trimming control is $\boldsymbol{u}_{\text{r0}}$,
then the linear model can be expressed as
\begin{equation}
\left\{ \begin{array}{l}
\Delta{\mathbf{{\dot{x}}}}_{\text{r}}\left(t\right)=\mathbf{A}_{\text{r}}\cdot\Delta{\mathbf{x}_{\text{r}}\left(t\right)}+\mathbf{B}_{\text{r}}\cdot\Delta\boldsymbol{u}_{\text{r}}\left(t\right)+\mathbf{G}_{\text{r}}\cdot{\mathbf{F}_{\text{r}}\left(t\right)}\\
\Delta\mathbf{p}{_{\text{pr}}}\left(t\right)=\mathbf{C}_{\text{r}}\cdot\Delta{\mathbf{x}_{\text{r}}}\left(t\right)
\end{array}\right.\label{Eq-rec1}
\end{equation}
where $\Delta{\mathbf{x}_{\text{r}}\triangleq\mathbf{x}_{\text{r}}-}\mathbf{x}_{\text{r0}}$
is the state vector of the linearized system, $\Delta\boldsymbol{u}_{\text{r}}\triangleq\boldsymbol{u}_{\text{r}}-\boldsymbol{u}_{\text{r0}}$
is the linearized control input vector and $\Delta\mathbf{p}${$_{\text{pr}}$}$\triangleq\mathbf{p}${$_{\text{pr}}$}$-\mathbf{p}{_{\text{pr0}}}$
is the probe position offset from the initial probe position $\mathbf{p}{_{\text{pr0}}}$.

\subsection{Control System}

\subsubsection{Autopilot}

Based on the linear model (\ref{Eq-rec1}), the autopilot can be simplified
as a state feedback controller \cite{lee2013estimation} in the form
as
\begin{eqnarray}
\Delta\boldsymbol{u}_{\text{r}}\left(t\right) & = & -\mathbf{K}_{\text{P}}\cdot\Delta{\mathbf{x}_{\text{r}}\left(t\right)}-\mathbf{K}_{\text{I}}\cdot\mathbf{e}_{\text{I}}\left(t\right)\label{Eq-rec2}\\
\mathbf{\dot{e}}_{\text{I}}\left(t\right) & = & \mathbf{p}{_{\text{pr}}\left(t\right)}-\mathbf{\hat{u}}_{\text{pr}}\left(t\right)\label{Eq-rec3}
\end{eqnarray}
where $\mathbf{\hat{u}}_{\text{pr}}\left(t\right)\in
\mathbb{R}
^{3}$ is the reference trajectory vector of the probe, $\mathbf{K}_{\text{P}}$
and $\mathbf{K}_{\text{I}}$ are the gain matrices. Essentially, Eq.~(\ref{Eq-rec2})
is a PI controller, where $-\mathbf{K}_{\text{P}}\cdot\Delta{\mathbf{x}_{\text{r}}}\left(t\right)$
is the state feedback control term for stabilizing the aircraft, and
${-}\mathbf{K}_{\text{I}}\cdot\mathbf{e}_{\text{I}}\left(t\right)$
is the integral control term for tracking the given trajectory. Since
it is very convenient to obtain $\mathbf{K}_{\text{P}}$ and $\mathbf{K}_{\text{I}}$
through LQR function in MATLAB, the procedures are omitted here. In
practice, a saturation function is required for $\mathbf{e}_{\text{I}}\left(t\right)$
in Eq.~(\ref{Eq-rec2}) to slow down the response speed and resist
integral saturation. For instance, the approaching speed should be
constrained within a reasonable range about 0.5m/s$\sim$1m/s, because
the probe should have enough closure speed to open the valve on the
drogue safely \cite{Dibley-2007-2}.

As analyzed in \cite{lee2013estimation,AirContrl}, when the autopilot
(\ref{Eq-rec2}) is well designed and the disturbance force {$\mathbf{F}$}${_{\text{r}}\left(t\right)}\equiv\mathbf{0}$,
the tracking error can converge to zero
\begin{equation}
\mathbf{\hat{u}}_{\text{pr}}\left(t\right)-\mathbf{p}{_{\text{pr}}}\left(t\right)\rightarrow\mathbf{0},\text{ as }t\rightarrow\infty.\label{Eq-rec2-1}
\end{equation}
However, in practice, the disturbance force {$\mathbf{F}$}${_{\text{r}}\left(t\right)}\neq\mathbf{0}$
and the terminal time $T\ll\infty$, then the tracking error cannot
reach zero at terminal time $T$. Therefore, an error term should
be added to Eq.~(\ref{Eq-rec2-1}) at $T$ as
\begin{equation}
\mathbf{\hat{u}}_{\text{pr}}\left(T\right)-\mathbf{p}{_{\text{pr}}}\left(T\right)=\mathbf{v}_{\text{pr}}\label{Eq-rec5}
\end{equation}
where $\mathbf{v}_{\text{pr}}\in
\mathbb{R}
^{3}$ is a bounded random disturbance term with $\left\Vert \mathbf{v}_{\text{pr}}\right\Vert \leq B_{\text{pr}}$.
The random disturbance $\mathbf{v}_{\text{pr}}$ may come from the
unrepeatable disturbances such as atmospheric turbulence.

\subsubsection{Objective of Docking Control}

\begin{figure}[tbh]
\centering \includegraphics[width=0.45\textwidth]{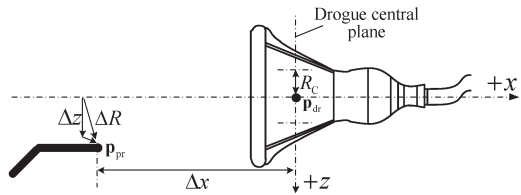}
\caption{Success and failure criteria of a docking attempt \cite{Dibley-2007-2}.}
\label{Fig_CaptureMissCriterion}
\end{figure}

According to \cite{Dibley-2007-2}, in each docking attempt, the receiver
should follow the drogue for seconds until the hose-drogue levels
off. Then, the receiver starts to drive the probe to approach the
drogue with a slow constant speed, until the probe hits the central
plane of the drogue as shown in Fig.~\ref{Fig_CaptureMissCriterion}.
The basic requirement for the AAR system is that the relative position
between the probe and the drogue (represented by the docking error
$\Delta{\mathbf{p}}_{\text{dr/pr}}$) can reach zero at the terminal
time $T$. In practice, the radial error $\Delta R_{\text{dr/pr}}\in
\mathbb{R}
_{+}$ is an important evaluation index for the docking performance which
defined in \textit{oyz} plane as
\begin{equation}
\Delta R_{\text{dr/pr}}\left(t\right)\triangleq\sqrt{\Delta y_{\text{dr/pr}}^{2}\left(t\right)+\Delta z_{\text{dr/pr}}^{2}\left(t\right)}.
\end{equation}
Since the docking error is inevitable due to disturbances, a threshold
radius (criterion radius) $R_{\text{C}}\in
\mathbb{R}
_{+}$ should be defined as
\begin{equation}
\Delta R_{\text{dr/pr}}\left(T\right)<R_{\text{C}}.\label{Eq-LearnCtrl0}
\end{equation}
If criterion (\ref{Eq-LearnCtrl0}) is satisfied, a success docking
is declared for this docking attempt \cite{Dibley-2007-2}. Otherwise,
a failure or miss is declared. In fact, according to the previous
definition, there is $\Delta x_{\text{dr/pr}}\left(T\right)\equiv0$.
Therefore, the terminal radial error $\Delta R_{\text{dr/pr}}\left(T\right)$
always equals to the terminal docking error $\Delta{\mathbf{p}}_{\text{dr/pr}}\left(T\right)$.

\section{TILC Design}

\label{Sec-3}

As shown in Fig.\ \ref{Fig-InnerCtrl}, the role of the TILC controller
in AAR system is the same as the human pilot in manned aerial refueling
system. The inputs of the TILC controller are the historical terminal
positions of the probe ${\mathbf{p}}_{\text{pr}}\left(T\right)$ and
the drogue ${\mathbf{p}}_{\text{dr}}\left(T\right)$, and the output
is the reference tracking trajectory $\mathbf{\hat{u}}_{\text{pr}}\left(t\right)$
which is further sent to the autopilot.

\subsection{TILC Controller}

The docking errors of the AAR system are mainly caused by two factors:
the drogue offset caused by the bow wave effect as described in Eq.~(\ref{Eq-wbow0});
and the tracking error caused by the response lag of the receiver
as described in Eq.~(\ref{Eq-rec5}). In order to compensate for
these docking errors, a simple and safe control strategy is letting
the probe always aims at a predicted fixed position $\mathbf{\hat{u}}_{\text{pr}}^{\left(k\right)}\left(t\right)\equiv\mathbf{\hat{u}}_{\text{pr}}^{\left(k\right)}$
during the docking stage. The predicted position $\mathbf{\hat{u}}_{\text{pr}}^{\left(k\right)}$
for the autopilot should have the following form
\begin{equation}
\mathbf{\hat{u}}_{\text{pr}}^{\left(k\right)}={\mathbf{p}}_{\text{dr}}^{\text{e0,}\left(k\right)}+\mathbf{u}_{\text{de,dr}}^{\left(k\right)}+\mathbf{u}_{\text{e,pr}}^{\left(k\right)}\label{Eq-Stg1-0}
\end{equation}
where ${\mathbf{p}}_{\text{dr}}^{\text{e0,}\left(k\right)}\in
\mathbb{R}
^{3}$ is the original stable position of the drogue, $\mathbf{u}_{\text{de,dr}}^{\left(k\right)}\in
\mathbb{R}
^{3}$ is an estimation term for the drogue position offset, and $\mathbf{u}_{\text{e,pr}}^{\left(k\right)}\in
\mathbb{R}
^{3}$ is an ILC term to compensate for the tracking error of the probe.
Note that, since ${\mathbf{p}}_{\text{dr}}^{\text{e0,}\left(k\right)}$
can be directly measured during the flight, it is treated as a known
parameter here. Then, $\mathbf{u}_{\text{de,dr}}^{\left(k\right)}$
and $\mathbf{u}_{\text{e,pr}}^{\left(k\right)}$ should be updated
in each iteration, and the updating laws are given below.

(1) the updating law of $\mathbf{u}_{\text{de,dr}}^{\left(k\right)}$
is given by
\begin{equation}
\mathbf{u}_{\text{de,dr}}^{\left(k\right)}=\mathbf{K}_{\alpha}\cdot\mathbf{u}_{\text{de,dr}}^{\left(k-1\right)}+\left(\mathbf{I}-\mathbf{K}_{\alpha}\right)\cdot\Delta{\mathbf{p}}_{\text{dr}}^{\text{e,}\left(k-1\right)}\label{Eq-Stg2-0}
\end{equation}
where $\mathbf{K}_{\alpha}=\text{diag}\left(k_{\alpha_{1}},k_{\alpha_{2}},k_{\alpha_{3}}\right)$
with $k_{\alpha_{1}},k_{\alpha_{2}},k_{\alpha_{3}}\in\left(0,1\right)$
is a constant diagonal matrix, and $\Delta{\mathbf{p}}_{\text{dr}}^{\text{e}}$
is the drogue terminal offset position as defined in Eq.~(\ref{Eq-wbow0})
whose iterative feature can be written as
\begin{equation}
\Delta{\mathbf{p}}_{\text{dr}}^{\text{e,}\left(k\right)}\triangleq\mathbf{p}{_{\text{dr}}^{(k)}}\left(T^{\left(k\right)}\right)-{\mathbf{p}}_{\text{dr}}^{\text{e0,}\left(k\right)}.\label{Eq-Stg2-1}
\end{equation}

(2) the updating law of $\mathbf{u}_{\text{e,pr}}^{\left(k\right)}$
is given by
\begin{equation}
\mathbf{u}_{\text{e,pr}}^{\left(k\right)}=\mathbf{u}_{\text{e,pr}}^{\left(k-1\right)}+\mathbf{K}_{p}\cdot\mathbf{e}_{\text{pr}}^{(k-1)}\label{Eq-Stg2-3}
\end{equation}
where $\mathbf{K}_{p}=\text{diag}\left(k_{p_{1}},k_{p_{2}},k_{p_{3}}\right)$
is a constant diagonal matrices with $k_{p_{1}},k_{p_{2}},k_{p_{3}}\in\left(0,1\right)$
and $\mathbf{e}_{\text{pr}}$ represents the probe terminal tracking
error with the $k^{\text{th}}$ iterative feature defined as
\begin{equation}
\mathbf{e}_{\text{pr}}^{(k)}\triangleq{\mathbf{p}}_{\text{dr}}^{\text{e0,}\left(k\right)}+\mathbf{u}_{\text{de,dr}}^{\left(k\right)}-\mathbf{p}{_{\text{pr}}^{(k)}}\left(T^{\left(k\right)}\right).\label{Eq-Stg2-4}
\end{equation}

\subsection{Convergence Analysis }

The following theorem provides the convergence condition under which
one can conclude the convergence property of the designed TILC controller
in Eq.\ (\ref{Eq-Stg1-0}).

\textbf{Theorem 1}. Consider the AAR system described by Eqs.~(\ref{Eq_hoseDorgueDyn})(\ref{Eq-rec})(\ref{Eq-rec2})
with the structure shown in Fig.~\ref{Fig-InnerCtrl}. Suppose (i)
the autopilot of the receiver aircraft in Eq.~(\ref{Eq-rec2}) is
well designed, and the probe terminal position satisfies Eq.~(\ref{Eq-rec5});
(ii) the TILC controller is designed as Eq.~(\ref{Eq-Stg1-0}), and
its parameters satisfy
\begin{equation}
0\leq k_{\alpha_{i}}<1,\text{ }0<k_{p_{i}}\leq1,\text{ }i=1,2,3.\label{Eq-Stg2-5}
\end{equation}
Then, through the repetitive docking attempts, the docking error $\Delta{\mathbf{p}}_{\text{dr/pr}}^{\left(k\right)}\left(T^{\left(k\right)}\right)$
will converge to a bound
\begin{equation}
\lim_{k\rightarrow\infty}\left\Vert \Delta{\mathbf{p}}_{\text{dr/pr}}^{\left(k\right)}\left(T^{\left(k\right)}\right)\right\Vert \leq B_{\text{dr/pr}}\label{Eq-Stg2-6}
\end{equation}
where
\begin{equation}
B_{\text{dr/pr}}=2\sqrt{B_{\text{pr}}^{2}+B_{\text{dr}}^{2}}\label{Eq-Stg2-7}
\end{equation}
in which $B_{\text{dr}}$ is the random disturbance bound of the drogue
position fluctuation as defined in Eq.~(\ref{Eq_hoseDrDyn2-3}) and
$B_{\text{pr}}$ is the random disturbance bound of the probe tracking
error as defined Eq.~(\ref{Eq-rec5}). In particular, if the random
disturbances are negligible, i.e., $B_{\text{dr}}=0$, $B_{\text{pr}}=0$,
then the docking error will converge to zero as
\begin{equation}
\left\Vert \Delta{\mathbf{p}}_{\text{dr/pr}}^{\left(k\right)}\left(T^{\left(k\right)}\right)\right\Vert \rightarrow0\text{, as }k\rightarrow\infty.\label{Eq-Stg2-8}
\end{equation}

\textbf{Proof.} See \textit{Appendix A}. $\square$

\subsection{Discussion}

Essentially, the term $\mathbf{u}_{\text{de,dr}}^{\left(k\right)}$
works as a low-pass filter, which is expected to provide a smooth
and robust estimation of the drogue offset caused by disturbances.
Then, with this term in $\mathbf{\hat{u}}_{\text{pr}}^{\left(k\right)}$,
the drogue offset can be compensated. The low-pass filter is adopted
instead of using the drogue offset position directly, which is because
the drogue is sensitive to disturbances.

The initial value for the proposed TILC method in Eq.\ (\ref{Eq-Stg1-0})
should be set to zero ($\mathbf{u}_{\text{de,dr}}^{\left(0\right)}=\mathbf{0}$,
$\mathbf{u}_{\text{e,pr}}^{\left(0\right)}=\mathbf{0}$) when there
is no historical learning data. In practice, $\mathbf{u}_{\text{de,dr}}^{\left(0\right)}$
has physical significance, namely the drogue position offset caused
by the receiver forebody flow field. Therefore, the initial value
for $\mathbf{u}_{\text{de,dr}}^{\left(0\right)}$ can be estimated
according to the historical learning data, the experience of human
pilots, or the calculation result from the hose-drogue model \cite{hose-link-model}
and the bow wave effect model \cite{dai2016modeling}. With the pre-estimated
initial value, the iteration speed of the proposed TILC method can
be improved.

Unlike other conventional ILC methods, the proposed TILC method does
not require the exact value of the terminal time $T$ and does not
require $T$ to be the same between iterations. It only requires the
terminal positions of the drogue and the probe, which is practical
for an actual AAR system.

\section{Simulation and Verification}

\label{Sec-4}

\subsection{Simulation Configuration}

A MATLAB/SIMULINK-based simulation environment has been developed
to simulate the docking stage of the AAR. The detailed introduction
of the modeling methods and the simulation parameters can be found
in the authors' previous work \cite{dai2016modeling}. A video has
also been released to introduce the AAR simulation environment and
demonstrate the TILC simulation results. The URL of the video is \url{https://youtu.be/VoplDA6D5fA}.

\subsection{TILC Simulation Results}

\subsubsection{Iterative Learning Process}

\begin{figure}[tbh]
\centering \includegraphics[width=0.45\textwidth]{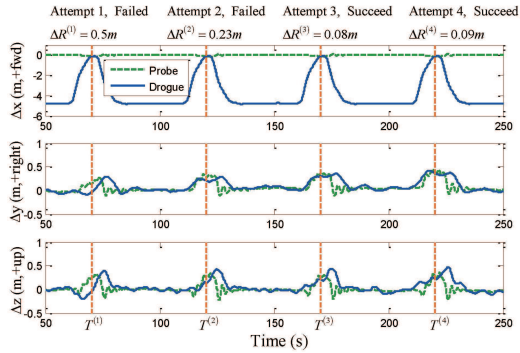}
\caption{Learning process with the proposed TILC method.}
\label{Fig_LearnProc}
\end{figure}

In order to verify the effectiveness of the proposed TILC method,
all the initial values in Eq.~(\ref{Eq-Stg1-0}) are set zeroes as
$\mathbf{u}_{\text{de,dr}}^{\left(0\right)}=\mathbf{0}$, $\mathbf{u}_{\text{e,pr}}^{\left(0\right)}=\mathbf{0}$,
and the learning procedures are shown in Fig.~\ref{Fig_LearnProc}.

In Fig.~\ref{Fig_LearnProc}, there are four docking attempts performed
in sequence (the four docking attempts start at time 50s, 100s, 150s
and 200s respectively), where the first two docking attempts fail,
and the following two attempts both succeed. In each attempt, the
probe moves close to until contact with the drogue at $T^{\left(k\right)}$
(marked by the vertical dotted lines), then the probe returns to the
standby postilion and gets ready for the next docking attempt.

In the first docking attempt as shown in Fig.~\ref{Fig_LearnProc},
the receiver remains at the standby position (5m behind the drogue,
with simulation time from 50s to 60s) to observe the drogue movement
and estimate the equilibrium position of the drogue. Then, the receiver
approaches the drogue to perform a docking attempt during the simulation
time from 60s to 71s in Fig.~\ref{Fig_LearnProc}. The docking control
ends at the terminal time $T^{\left(1\right)}=71s$, and this docking
attempt is declared as a failure because the radial error $\Delta R_{\text{dr/pr}}^{\left(1\right)}=0.5m$
is larger than the desired radial error threshold $R_{C}=0.15\text{m}$.

With more docking attempts (not presented in Fig.\ \ref{Fig_LearnProc})
are simulated, a docking success rate over 90\% will be obtained under
the given threshold $R_{C}=0.15\text{m}$. According to the Monte
Carlo simulations, the success rate depends on many factors including
the docking error threshold $R_{C}$, the strength of the atmospheric
turbulence, and other random disturbances. The simulation results
are consistent with results in \cite{Dibley-2007-2}\cite{Bhandari-2013-8}.
When the aerodynamic disturbances are strong, both the drogue position
oscillation and the receiver tracking error will be significant, then
the success rate will be low.

\subsubsection{Aerodynamic Disturbance Simulations}

\begin{figure}[tbh]
\centering \includegraphics[width=0.45\textwidth]{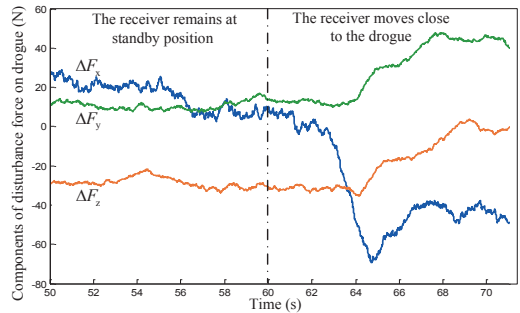}
\caption{Total aerodynamic disturbing force applied on the drogue.}
\label{Fig_totalAeroDisOnDro}
\end{figure}

Fig.~\ref{Fig_totalAeroDisOnDro} presents the total aerodynamic
disturbance force $\mathbf{F}_{\text{total}}=\left[\Delta F_{\text{x}},\Delta F_{\text{y}},\Delta F_{\text{z}}\right]^{\text{T}}$
applied on the drogue during the first docking attempt (50s$\sim$71s)
in Fig.\ \ref{Fig_LearnProc}. In this simulation, the tanker vortex
disturbance comes from the model presented in \cite{Vortex-1}, the
wind gust and the atmospheric turbulence come from the MATLAB/SIMULINK
Aerospace Blockset based on the mathematical representations from
Military Specification MIL-F-8785C, and the bow wave effect disturbance
comes from the authors' previous work \cite{wei2016drogue}. When
the receiver remains at the standby position (50s$\sim$60s in Fig.~\ref{Fig_totalAeroDisOnDro}),
the drogue is far away from the receiver and the disturbance forces
mainly come from the tanker vortex and the atmospheric turbulence
as illustrated on the left half of Fig.~\ref{Fig_totalAeroDisOnDro}.
As the receiver moves closer to the drogue, the receiver bow wave
starts to cause a large disturbance force on the drogue as illustrated
on the right half of Fig.~\ref{Fig_totalAeroDisOnDro}.

\begin{figure}[tbh]
\centering \includegraphics[width=0.45\textwidth]{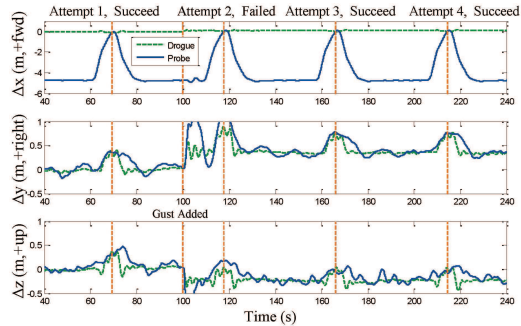}
\caption{Simulation with the initial value and the wind gust disturbance.}
\label{Fig_compSim}
\end{figure}

A comprehensive simulation is performed to verify the performance
of the proposed TILC method with the initial value from the previous
learning results. In addition to the atmospheric turbulence and the
bow wave disturbance as shown in Fig.~\ref{Fig_totalAeroDisOnDro},
a wind gust ($5m/s$ in the lateral direction and vertical direction
respectively) is added at 100s to verify the control effect of the
proposed method under aerodynamic disturbances. The simulation results
are presented in Fig.~\ref{Fig_compSim}.

It can be observed from Fig.~\ref{Fig_compSim} that, with a good
initial value, the docking control succeeds at the first attempt.
Then, the second docking attempt (115s in Fig.~\ref{Fig_compSim})
fails due to the addition of a strong wind gust at 100s. In the next
two docking attempts (165s and 215s in Fig.~\ref{Fig_compSim}),
the controller can rapidly recover and achieve successful docking
control without being much affected by the wind gust disturbance.
The simulation results demonstrate that the proposed TILC method has
a certain ability to resist the aerodynamic disturbances.

\section{Conclusions}

\label{Sec-5}

This paper studies the model of the probe-drogue aerial refueling
system under aerodynamic disturbances, and proposes a docking control
method based on terminal iterative learning control to compensate
for the docking errors caused by aerodynamic disturbances. The designed
controller works as an additional unit for the trajectory generation
function of the original autopilot system. Simulations based on our
previously published simulation environment show that the proposed
control method has a fast learning speed to achieve a successful docking
control under aerodynamic disturbances including the bow wave effect.

\section*{Acknowledgments}

This work was supported by the National Key Project of Research and
Development Plan under Grant 2016YFC1402500 and the National Natural
Science Foundation of China under Grant 61473012.

\section*{Appendices}

\subsection{Proof of Theorem 1}

\label{App-1}First, define the $\mathbf{p}{_{\text{pr}}^{\left(k\right)}}\left(T^{\left(k\right)}\right)$
as the probe terminal position in the $k^{\text{th}}$ docking attempt.
Then, according to Eq.~(\ref{Eq-rec5}), one has
\begin{equation}
\mathbf{p}{_{\text{pr}}^{\left(k\right)}}\left(T^{\left(k\right)}\right)=\mathbf{\hat{u}}_{\text{pr}}^{\left(k\right)}-\mathbf{v}_{\text{pr}}^{\left(k\right)}\label{Eq-convAna1}
\end{equation}
where, $\mathbf{\hat{u}}_{\text{pr}}^{\left(k\right)}$ can be further
expressed by Eq.~(\ref{Eq-Stg1-0}), which yields
\begin{equation}
{\mathbf{p}}_{\text{dr}}^{\text{e0,}\left(k\right)}+\mathbf{u}_{\text{de,dr}}^{\left(k\right)}-\mathbf{p}{_{\text{pr}}^{\left(k\right)}}\left(T^{\left(k\right)}\right)=\mathbf{v}_{\text{pr}}^{\left(k\right)}-\mathbf{u}_{\text{e,pr}}^{\left(k\right)}.\label{Eq-convAna2}
\end{equation}
Meanwhile, according to the definition of $\mathbf{e}_{\text{pr}}^{(k)}$
in Eq.~(\ref{Eq-Stg2-4}), one has
\begin{equation}
\mathbf{e}_{\text{pr}}^{(k)}=\mathbf{v}_{\text{pr}}^{\left(k\right)}-\mathbf{u}_{\text{e,pr}}^{\left(k\right)}.\label{Eq-convAna3}
\end{equation}
Thus, substituting Eq.\ (\ref{Eq-Stg2-3}) into Eq.~(\ref{Eq-convAna3})
gives
\begin{equation}
\mathbf{e}_{\text{pr}}^{(k)}=\left(\mathbf{I}-\mathbf{K}_{p}\right)\cdot\mathbf{e}_{\text{pr}}^{(k-1)}+\mathbf{\tilde{v}}_{\text{pr}}^{\left(k-1\right)}\label{Eq-convAna6}
\end{equation}
where
\begin{equation}
\mathbf{\tilde{v}}_{\text{pr}}^{\left(k-1\right)}\triangleq\mathbf{v}_{\text{pr}}^{\left(k\right)}-\mathbf{v}_{\text{pr}}^{\left(k-1\right)}.\label{eq:Vpr}
\end{equation}

Second, according to Eq.~(\ref{Eq-wbow0}), the drogue terminal position
$\mathbf{p}{_{\text{dr}}^{(k)}}\left(T^{\left(k\right)}\right)$ in
the $k^{\text{th}}$ docking attempt is given by
\begin{equation}
\mathbf{p}{_{\text{dr}}^{(k)}}\left(T^{\left(k\right)}\right)={\mathbf{p}}_{\text{dr}}^{\text{e0,}\left(k\right)}+\Delta{\mathbf{p}}_{\text{dr}}^{\text{e,}\left(k\right)}\label{Eq-convAna8}
\end{equation}
where ${\mathbf{p}}_{\text{dr}}^{\text{e0,}\left(k\right)}$ is the
drogue original equilibrium position, and $\Delta{\mathbf{p}}_{\text{dr}}^{\text{e,}\left(k\right)}$
is the terminal position offset. According to Eq.~(\ref{Eq_hoseDrDyn2-3}),
$\Delta{\mathbf{p}}_{\text{dr}}^{\text{e,}\left(k\right)}$ comes
from the bow wave effect and can be expressed

\begin{equation}
\Delta{\mathbf{p}}_{\text{dr}}^{\text{e,}\left(k\right)}=\mathbf{m}_{0}+\mathbf{M}_{1}\cdot\Delta{\mathbf{p}}_{\text{dr/pr}}^{\left(k\right)}\left(T^{\left(k\right)}\right)+\mathbf{v}_{\text{dr}}^{\left(k\right)}.\label{Eq-convAna7-1}
\end{equation}
Thus, the docking error along the iteration axis is given by
\begin{equation}
\Delta{\mathbf{p}}_{\text{dr/pr}}^{\left(k\right)}\left(T^{\left(k\right)}\right)=\mathbf{p}{_{\text{dr}}^{\left(k\right)}}\left(T^{\left(k\right)}\right)-\mathbf{p}{_{\text{pr}}^{\left(k\right)}}\left(T^{\left(k\right)}\right).\label{Eq-convAna9}
\end{equation}
Substituting Eqs.~(\ref{Eq-convAna8})(\ref{Eq-convAna7-1})(\ref{Eq-convAna9})
into Eqs.~(\ref{Eq-Stg2-0})(\ref{Eq-Stg2-1}) gives
\begin{equation}
\Delta{\mathbf{p}}_{\text{dr/pr}}^{\left(k\right)}\left(T^{\left(k\right)}\right)=\mathbf{A}_{1}\cdot\Delta{\mathbf{p}}_{\text{dr/pr}}^{\left(k-1\right)}\left(T^{\left(k-1\right)}\right)+\mathbf{A}_{2}\cdot\mathbf{e}_{\text{pr}}^{(k-1)}+\mathbf{\tilde{v}}_{\text{dr}}^{\left(k-1\right)}\label{Eq-convAna10}
\end{equation}
where
\begin{eqnarray}
\mathbf{A}_{1} & \triangleq & \left(\mathbf{M}_{1}-\mathbf{I}\right)^{-1}\left(\mathbf{M}_{1}-\mathbf{K}_{\alpha}\right)=\mathbf{I-}\left(\mathbf{I-M}_{1}\right)^{-1}\left(\mathbf{I}-\mathbf{K}_{\alpha}\right)\label{Eq-convAna11-1}\\
\mathbf{A}_{2} & \triangleq & \left(\mathbf{M}_{1}-\mathbf{I}\right)^{-1}\left(\mathbf{K}_{p}+\mathbf{K}_{\alpha}-\mathbf{I}\right)\label{Eq-convAna11-2}\\
\mathbf{\tilde{v}}_{\text{dr}}^{\left(k-1\right)} & \triangleq & \left(\mathbf{M}_{1}-\mathbf{I}\right)^{-1}\left(\mathbf{v}_{\text{dr}}^{\left(k-1\right)}-\mathbf{v}_{\text{dr}}^{\left(k\right)}\right).\label{Eq-convAna11-3}
\end{eqnarray}
For simplicity, an augmented system is defined as
\begin{equation}
\mathbf{X}^{\left(k\right)}=\mathbf{A\cdot X}^{\left(k-1\right)}+\mathbf{v}^{\left(k-1\right)}\label{Eq-convAna13}
\end{equation}
where

\begin{eqnarray}
\mathbf{X}^{\left(k\right)} & \triangleq & \left[\begin{array}{c}
\Delta{\mathbf{p}}_{\text{dr/pr}}^{\left(k\right)}\left(T^{\left(k\right)}\right)\\
\mathbf{e}_{\text{pr}}^{(k)}
\end{array}\right],\mathbf{v}^{\left(k\right)}\triangleq\left[\begin{array}{c}
\mathbf{\tilde{v}}_{\text{dr}}^{\left(k\right)}\\
\mathbf{\tilde{v}}_{\text{pr}}^{\left(k\right)}
\end{array}\right]\label{Eq-convAna12-0}\\
\mathbf{A} & \triangleq & \left[\begin{array}{cc}
\mathbf{A}_{1} & \mathbf{A}_{2}\\
\mathbf{0}_{3\times3} & \mathbf{A}_{3}
\end{array}\right],\;\mathbf{A}_{3}\triangleq\mathbf{I}-\mathbf{K}_{p}.\label{Eq-convAna12}
\end{eqnarray}
Furthermore, Eq.\ (\ref{Eq-convAna12-0}) can be written into the
following form
\begin{equation}
\mathbf{X}^{\left(k\right)}=\mathbf{A}^{k}\cdot\mathbf{X}^{\left(0\right)}+\sum_{i=0}^{k-1}\mathbf{A}^{i}\mathbf{v}^{\left(k-i\right)}.\label{eq:xkak}
\end{equation}

Since $\mathbf{M}_{1}$ is a negative definite matrix, according to
Eqs.\ (\ref{Eq-convAna11-1})(\ref{Eq-convAna11-1})(\ref{Eq-convAna12-0}),
it is easy to verify that the spectral radius of $\mathbf{A}$ is
smaller than 1 ($\rho\left(\mathbf{A}\right)<1$) when the following
constraint is satisfied
\begin{equation}
0\leq k_{\alpha_{i}}<1,\text{ }0<k_{p_{i}}\leq1\text{, }i=1,2,3.\label{Eq-convAna17-2}
\end{equation}
Moreover, since the disturbances $\mathbf{v}_{\text{pr}}^{\left(k\right)}$
and $\mathbf{v}_{\text{dr}}^{\left(k\right)}$ are both bounded with
$\left\Vert \mathbf{v}_{\text{pr}}^{\left(k\right)}\right\Vert \leq B_{\text{pr}}$
and $\left\Vert \mathbf{v}_{\text{dr}}^{\left(k\right)}\right\Vert \leq B_{\text{dr}}$,
it is easy to obtain from Eqs.\ (\ref{eq:Vpr})(\ref{Eq-convAna11-1})(\ref{Eq-convAna12-0})
that $\mathbf{v}^{\left(k\right)}$ is also bounded with
\begin{equation}
\left\Vert \mathbf{v}^{\left(k\right)}\right\Vert \leq2\sqrt{B_{\text{pr}}^{2}+B_{\text{dr}}^{2}}.\label{eq:vBr}
\end{equation}
Then, substituting Eq.\ (\ref{eq:vBr}) into Eq.\ (\ref{eq:xkak})
gives
\begin{equation}
\begin{array}{ll}
\left\Vert \mathbf{X}^{\left(k\right)}\right\Vert  & \leq\left\Vert \mathbf{A}\right\Vert ^{k}\left\Vert \mathbf{X}^{\left(0\right)}\right\Vert +\sum_{i=0}^{k-1}\left\Vert \mathbf{A}\right\Vert ^{i}\left\Vert \mathbf{v}^{\left(k-i\right)}\right\Vert \\
 & \leq\left\Vert \mathbf{A}\right\Vert ^{k}\left\Vert \mathbf{X}^{\left(0\right)}\right\Vert +2\sqrt{B_{\text{pr}}^{2}+B_{\text{dr}}^{2}}\sum_{i=0}^{k-1}\left\Vert \mathbf{A}\right\Vert ^{i}\\
 & =\left\Vert \mathbf{A}\right\Vert ^{k}\left\Vert \mathbf{X}^{\left(0\right)}\right\Vert +2\sqrt{B_{\text{pr}}^{2}+B_{\text{dr}}^{2}}\left(1-\left\Vert \mathbf{A}\right\Vert ^{k}\right).
\end{array}\label{eq:xka}
\end{equation}
When the constraint in Eq.\ (\ref{Eq-convAna17-2}) is satisfied,
one has
\begin{equation}
\rho\left(\mathbf{A}\right)<1\Rightarrow\lim_{k\rightarrow\infty}\left\Vert \mathbf{A}\right\Vert ^{k}=0
\end{equation}
which yields from Eq.\ (\ref{eq:xka}) that
\begin{equation}
\lim_{k\rightarrow\infty}\left\Vert \mathbf{X}^{\left(k\right)}\right\Vert \leq2\sqrt{B_{\text{pr}}^{2}+B_{\text{dr}}^{2}}.\label{eq:xk46}
\end{equation}

According to the definition of $\mathbf{X}^{\left(k\right)}$ in Eq.\ (\ref{Eq-convAna12-0}),
one has
\begin{equation}
\left\Vert \Delta{\mathbf{p}}_{\text{dr/pr}}^{\left(k\right)}\left(T^{\left(k\right)}\right)\right\Vert \leq\left\Vert \mathbf{X}^{\left(k\right)}\right\Vert .\label{eq:xk47}
\end{equation}
Combining Eq.\ (\ref{eq:xk46}) and (\ref{eq:xk47}) gives
\begin{equation}
\lim_{k\rightarrow\infty}\left\Vert \Delta{\mathbf{p}}_{\text{dr/pr}}^{\left(k\right)}\left(T^{\left(k\right)}\right)\right\Vert \leq2\sqrt{B_{\text{pr}}^{2}+B_{\text{dr}}^{2}}=B_{\text{dr/pr}}.\label{eq:kwuqiong}
\end{equation}
Thus, the docking error $\Delta{\mathbf{p}}_{\text{dr/pr}}^{\left(k\right)}\left(T^{\left(k\right)}\right)$
will converge to a bound $B_{\text{dr/pr}}$ as $k\rightarrow\infty$.
In particular, by substituting $B_{\text{dr}}=0\text{, }B_{\text{pr}}=0$
into Eq.\ (\ref{eq:kwuqiong}), one has $\lim_{k\rightarrow\infty}\left\Vert \Delta{\mathbf{p}}_{\text{dr/pr}}^{\left(k\right)}\left(T^{\left(k\right)}\right)\right\Vert =0$.



\section*{References}

 \bibliographystyle{aiaa}
\bibliography{TILC_AAR}

\begin{thebibliography}{10}
\newcommand{\enquote}[1]{``#1''}
\expandafter\ifx\csname urlstyle\endcsname\relax
  \providecommand{\doi}[1]{doi:\discretionary{}{}{}#1}\else
  \providecommand{\doi}{doi:\discretionary{}{}{}\begingroup
  \urlstyle{rm}\Url}\fi

\bibitem{Nalepka-2005-1}
Nalepka, J.~P. and Hinchman, J.~L., \enquote{Automated aerial refueling:
  extending the effectiveness of unmanned air vehicles,} in \enquote{AIAA
  Modeling and Simulation Technologies Conference and Exhibit,} AIAA Paper
  2005-6005, Aug. 2005, \newline\doi{10.2514/6.2005-6005}.

\bibitem{Dibley-2007-2}
Dibley, R.~P., Allen, M.~J., and Nabaa, N., \enquote{Autonomous Airborne
  Refueling Demonstration Phase I Flight-Test Results,} in \enquote{AIAA
  Atmospheric Flight Mechanics Conference and Exhibit,} AIAA Paper 2007-6639,
  Aug. 2007, \newline\doi{10.2514/6.2007-6639}.

\bibitem{AAR-2014}
Thomas, P.~R., Bhandari, U., Bullock, S., Richardson, T.~S., and Du~Bois,
  J.~L., \enquote{Advances in air to air refuelling,} \emph{Progress in
  Aerospace Sciences}, Vol.~71, 2014, pp. 14--35,
  \newline\doi{10.1016/j.paerosci.2014.07.001}.

\bibitem{Bhandari-2013-8}
Bhandari, U., Thomas, P.~R., Bullock, S., Richardson, T.~S., and du~Bois,
  J.~L., \enquote{Bow Wave Effect in Probe and Drogue Aerial Refuelling,} in
  \enquote{AIAA Guidance, Navigation, and Control Conference,} AIAA Paper
  2013-4695, Aug. 2013, \newline\doi{10.2514/6.2013-4695}.

\bibitem{dai2016modeling}
Dai, X., Wei, Z.-B., and Quan, Q., \enquote{Modeling and simulation of bow wave
  effect in probe and drogue aerial refueling,} \emph{Chinese Journal of
  Aeronautics}, Vol.~29, No.~2, 2016, pp. 448--461,
  \newline\doi{10.1016/j.cja.2016.02.001}.

\bibitem{wei2016drogue}
Wei, Z.-B., Dai, X., Quan, Q., and Cai, K.-Y., \enquote{Drogue dynamic model
  under bow wave in probe-and-drogue refueling,} \emph{IEEE Transactions on
  Aerospace and Electronic Systems}, Vol.~52, No.~4, 2016, pp. 1728--1742,
  \newline\doi{10.1109/TAES.2016.140912}.

\bibitem{tandale2006trajectory}
Tandale, M.~D., Bowers, R., and Valasek, J., \enquote{Trajectory tracking
  controller for vision-based probe and drogue autonomous aerial refueling,}
  \emph{Journal of Guidance, Control, and Dynamics}, Vol.~29, No.~4, 2006, pp.
  846--857, \newline\doi{10.2514/1.19694}.

\bibitem{zhu2016vision}
Zhu, H., Yuan, S., and Shen, Q., \enquote{Vision/GPS-based docking control for
  the UAV Autonomous Aerial Refueling,} in \enquote{Guidance, Navigation and
  Control Conference (CGNCC), 2016 IEEE Chinese,} IEEE, 2016, pp. 1211--1215,
  \newline\doi{10.1109/CGNCC.2016.7828960}.

\bibitem{liu2017modeling}
Liu, Z., Liu, J., and He, W., \enquote{Modeling and vibration control of a
  flexible aerial refueling hose with variable lengths and input constraint,}
  \emph{Automatica}, Vol.~77, 2017, pp. 302--310,
  \newline\doi{10.1016/j.automatica.2016.11.002}.

\bibitem{Vortex-1}
Dogan, A., Lewis, T.~A., and Blake, W., \enquote{Flight data analysis and
  simulation of wind effects during aerial refueling,} \emph{Journal of
  Aircraft}, Vol.~45, No.~6, 2008, pp. 2036--2048,
  \newline\doi{10.2514/1.36797}.

\bibitem{lee2013estimation}
Lee, J.~H., Sevil, H.~E., Dogan, A., and Hullender, D., \enquote{Estimation of
  receiver aircraft states and wind vectors in aerial refueling,} \emph{Journal
  of Guidance, Control, and Dynamics}, Vol.~37, No.~1, 2013, pp. 265--276,
  \newline\doi{10.2514/1.59783}.

\bibitem{Khan-2014-9}
Khan, O. and Masud, J., \enquote{Trajectory analysis of basket engagement
  during aerial refueling,} in \enquote{AIAA Atmospheric Flight Mechanics
  Conference,} AIAA Paper 2014-0190, Jan. 2014,
  \newline\doi{10.2514/6.2014-0190}.

\bibitem{bristow2006survey}
Bristow, D.~A., Tharayil, M., and Alleyne, A.~G., \enquote{A survey of
  iterative learning control,} \emph{IEEE Control Systems}, Vol.~26, No.~3,
  2006, pp. 96--114, \newline\doi{10.1109/MCS.2006.1636313}.

\bibitem{ahn2007iterative}
Ahn, H.-S., Chen, Y., and Moore, K.~L., \enquote{Iterative learning control:
  Brief survey and categorization,} \emph{IEEE Transactions on Systems, Man,
  and Cybernetics, Part C (Applications and Reviews)}, Vol.~37, No.~6, 2007,
  pp. 1099--1121, \newline\doi{10.1109/TSMCC.2007.905759}.

\bibitem{valasek2005vision}
Valasek, J., Gunnam, K., Kimmett, J., Junkins, J.~L., Hughes, D., and Tandale,
  M.~D., \enquote{Vision-based sensor and navigation system for autonomous air
  refueling,} \emph{Journal of Guidance, Control, and Dynamics}, Vol.~28,
  No.~5, 2005, pp. 979--989, \newline\doi{10.2514/1.11934}.

\bibitem{chen1999iterative}
Chen, Y. and Wen, C., \emph{Iterative learning control: convergence, robustness
  and applications}, Springer-Verlag, 1999, \newline\doi{10.1007/BFb0110114}.

\bibitem{chi2014improved}
Chi, R., Hou, Z., Jin, S., and Wang, D., \enquote{Improved data-driven optimal
  TILC using time-varying input signals,} \emph{Journal of Process Control},
  Vol.~24, No.~12, 2014, pp. 78--85,
  \newline\doi{10.1016/j.jprocont.2014.07.007}.

\bibitem{NATO-2004-3}
NATO, \enquote{ATP-56(B) Air-to-Air Refuelling,} Tech. rep., NATO, 2010.

\bibitem{hose-link-model}
Ro, K. and Kamman, J.~W., \enquote{Modeling and simulation of hose-paradrogue
  aerial refueling systems,} \emph{Journal of Guidance, Control, and Dynamics},
  Vol.~33, No.~1, 2010, pp. 53--63, \newline\doi{10.2514/1.45482}.

\bibitem{AirContrl}
Stevens, B.~L. and Lewis, F.~L., \emph{Aircraft Control and Simulation}, John
  Wiley \& Sons, 2004, \newline\doi{10.1108/aeat.2004.12776eae.001}.

\end{thebibliography}

\end{document}